\documentclass[a4paper,10pt]{article}
\usepackage[all]{xy} 
\usepackage{macros}
\title{On the power of quantum, one round, two prover interactive proof systems}

\author{
     Alex Rapaport and Amnon Ta-Shma\\
Department of Computer Science\\
Tel-Aviv University\\
Israel 69978.\\
email: rapapo,amnon@post.tau.ac.il.
\thanks{This research was supported by a united
states-Israel binational science foundation grant and by the EU
integrated project QAP.} }

\begin{document}


\maketitle

\begin{abstract}
We analyze quantum two prover one round interactive proof systems,
in which noninteracting provers can share unlimited entanglement.
The maximum acceptance probability is characterized as a
superoperator norm. We get some partial results about the
superoperator norm, and in particular we analyze the "rank one"
case.
\end{abstract}

\section{Introduction}
Classical interactive proof systems allow an interaction between an
efficient verifier and an all powerful prover. Classical interactive
proof systems are quite powerful: one prover can prove theorems in
PSPACE to an efficient verifier \cite{IP-PSPACE} while two or more
powerful provers that cannot interact between themselves can prove
the whole of NEXP \cite{Nexp2}.

Kitaev and Watrous \cite{KitWat} studied the power of interaction
between an efficient quantum verifier and a single prover. They
prove that such a proof system is at least as powerful as a
classical one prover proof system but not as powerful as classical
two provers ($\PSPACE \subseteq \QIP \subseteq \EXP$). Moreover they
show that in the quantum case $3$ communication messages are enough
($\QIP=\QIP(3)$). They also show how to achieve perfect completeness
and parallel amplification for the model.

The quantum multiprover case is more complicated. As in the
classical case the provers cannot interact between themselves. There
are three models concerning the initial state of the provers private
qubits. In one model they are not allowed to share any prior
entanglement at all, in the second they are allowed to share limited
entanglement and in the third they can share unlimited entanglement.
Kobayashi and Matsumoto \cite{QMIP} prove that without entanglement
quantum multiprover proofs are as powerful as classical.  They also
prove that if we limit prior entanglement to be polynomial in the
input size the power of the proof can only decrease.

In this paper we concentrate on the case of two quantum provers with
unlimited prior entanglement and one round of communication. The
power of such proofs is not known. On the one hand more entanglement
gives the provers power to prove more languages to the verifier, but
on the other hand it gives them more power to cheat the verifier.
The only prior result we are aware of, is that of Kempe and Vidick
\cite{Julia}, that such provers can prove NP with perfect
completeness and some non-negligible soundness, to a verifier whose
space is limited to be logarithmic in the input size.

The problem we are facing touches the basic question of what
entanglement can achieve, and how to quantify it. There are many
demonstrations of the power of entanglement (e.g., teleportation
\cite{Teleport} and  superdense coding \cite{DenseCode}). There is
also a natural measure for measuring the amount of entanglement in
\emph{pure states}  \cite{EntMes}. Yet, there is no good measure for
the amount of entanglement in \emph{mixed states}.

Another demonstration of the power of entanglement are nonlocal
games. In those games Alice and Bob play as provers against a fixed
verifier. Their goal is to make him accept. The value of the game is
the probability a verifier accepts when Alice and Bob play
optimally. Alice and Bob cannot interact during the game but in the
quantum model they may share prior entanglement. The CHSH and the
Magic Square games are two examples presented in \cite{Str} and
\cite{Aravind} for games in which quantum provers outperform the
classical provers and violate Bell inequalities for classical
correlation between noninteracting parties. In the case of the Magic
Square game there is even a perfect quantum strategy that achieves
game value $1$. The problem we work on is a  strong generalization
of quantum nonlocal games.

It is fair to say that entanglement is far from being understood.
In particular, we don't even understand
whether infinite entanglement gives additional power over limited
entanglement, and this is the core of the problem we try to deal
with in this work.

Our approach is to generalize the direction Watrous and Kitaev
\cite{KitWat} took with the quantum \emph{single} prover case. They
gave an algebraic characterization for the maximum acceptance
probability of a fixed verifier in terms of the diamond
superoperator norm. Then they used a nice algebraic property of the
diamond norm, proved previously by Kitaev \cite{kitaev2002caq}, to
get strong results about quantum single prover proofs.

We manage to get an algebraic characterization of one-round,
two-prover games. We define a "product superoperator norm" and use
it to characterize the maximum acceptance probability of a fixed
verifier in the quantum two prover, one round case. However, we are
unable to analyze it algebraically. We get some partial results and
in particular we analyze the "rank one" case. Even this case is
nontrivial. We also present some hypotheses about our
characterization and give their implications on the power of the
proof system.

\section{Preliminaries and Background}

\subsection{Basic Notation}

For a Hilbert space $\hs{H}$ with dimension $\dim(\hs{H})$ we denote
by $L(\hs{H})$ the set of all linear operators over $\hs{H}$ and by
$U(\hs{H})$ the set of all unitary operators over $\hs{H}$.
$I_{\hs{H}}$ denotes the identity operator over $\hs{H}$. A
\emph{superoperator} $T:L(\hs{H}_1) \rightarrow L(\hs{H}_2)$ is a
linear mapping from $L(\hs{H}_1)$ to $L(\hs{H}_2)$.

\begin{deff} The \emph{trace out} operator is a superoperator from $L(\hs{H}_1
\tensor \hs{H}_2)$ to $L(\hs{H}_1)$ defined by
\[
\Tr_{\hs{H}_2}(A \tensor B) = \Tr(B) \cdot A
\]
for $A \in L(\hs{H}_1)$ and $B \in L(\hs{H}_2)$ and extended
linearly to all of $L(\hs{H}_1 \tensor \hs{H}_2)$.
\end{deff}

It can be checked that for $X \in L(\hs{H}_1 \tensor \hs{H}_2)$,
 $\Tr_{\hs{H}_2}(X)$ is independent of the representation
$X=\sum_i{A_i \tensor B_i}$. Also it is easy to check that

\begin{equation}
\Tr(\Tr_{\hs{H}_2}(X))=\Tr(X)
\end{equation}
 and that
\begin{equation}
\Tr_{\hs{H}_2}((C\tensor I)X)=C\Tr_{\hs{H}_2}(X)
\end{equation}
for any $C \in L(\hs{H}_1)$.

\subsection{Quantum Interactive Proof Systems}

In quantum interactive proof systems the verifier and the provers
are quantum players. The protocol lives in $\hs{V} \tensor \hs{M}_1
\tensor \cdots \tensor \hs{M}_k \tensor \hs{P}_1 \tensor \cdots
\tensor \hs{P}_k$ where $\hs{V}$ is the verifier private register,
  $\hs{M}_i$ is the message register between the verifier and the $i$'th prover
 and $\hs{P}_i$ is the $i$'th prover private register. $\hs{V}$ and
$\hs{M}_i$ are of size polynomial in the input length. In every
round of the proof the verifier applies a unitary transformation on
$\hs{V} \tensor \hs{M}_1 \tensor \cdots \tensor \hs{M}_k$ after
which the $\hs{M}_i$ register is sent to the $i$'th prover who
applies a unitary transformation on $\hs{M}_i \tensor \hs{P}_i$ and
sends $\hs{M}_i$ back to the verifier. Because of the safe storage
and the locality principle it is convenient to assume without loss of generality
that there is only one measurement done by the verifier at the end, based on which
he accepts or rejects.

$\QIP(m)$ (Quantum \IP) is the class of languages that can be proved
to a quantum verifier with $c=\frac{2}{3}$ and $s=\frac{1}{3}$ by a
single quantum prover with at most $m$ messages passed between the
prover and the verifier. Note that in the quantum model we usually
count the actual number of passed messages in each direction and not
the number of rounds, as is customary in the classical model.

Kitaev and Watrous \cite{KitWat} proved that $\PSPACE \subseteq
\QIP=\QIP(3)\subseteq \EXP$. There is no similar result in classical
\IP. They also showed that any language in $\QIP(3)$ has a proof with perfect completeness.
 Also $\QIP(3)$ has perfect parallel amplification.


We now turn to \emph{multiprover proof systems}. An important
parameter of multiprover quantum interactive proof systems is the
maximal amount of entangled qubits the provers are allowed to share
(if at all) in the initial state of $\hs{P}_1 \tensor \cdots \tensor
\hs{P}_k$. We say that the provers have
\emph{$q(\abs{x})$-prior-entanglement} if all the provers hold at most
$q(\abs{x})$ entangled qubits in the initial state.

\begin{definition} Fix functions $k(\abs{x}),m(\abs{x}),q(\abs{x})\geq
0$. $\QMIP(k,m,q)$ is the class of languages $L$ for which there is
an interactive proof system with
\begin{itemize}
  \item
  $k$ quantum provers.
  \item
  $m$ communication rounds.
  \item
  The initial state $\ket{\psi}$, between the provers is $q(\abs{x})$-prior-entangled.
\end{itemize}
such that
\begin{enumerate}
  \item If $x \in L$ then there exist quantum provers $P_1,\ldots,P_k$ and $\ket{\psi}$  for which
  $V_x$ accepts with probability at least $\frac{2}{3}$.
  \item If $x \not \in L$ then for all quantum provers $P_1,\ldots,P_k$ and $\ket{\psi}$, $V_x$
  accepts with probability at most $\frac{1}{3}$.
\end{enumerate}
\end{definition}

Note that we define $m$ as the number of communication rounds, and
not as the number of communication messages. Since we study only the
case of one round two messages, the classical convention is more
appropriate in this case.

Denote
\begin{eqnarray*}
\QMIP(k,m) &=& \QMIP(k,m,0)\\
\QMIP^{poly}(k,m) &=& \QMIP(k,m,poly)\\
\QMIP^*(k,m) &=& \QMIP(k,m,\infty)
\end{eqnarray*}
Kobayashi and Matsumoto prove in \cite{QMIP} that
$$\QMIP(poly,poly)=\MIP(poly,poly)=\NEXP$$
Also, they proved that if the provers have
$poly(\abs{x})$-prior-entanglement then we can assume that
$\dim(\hs{P}_i)=2^{poly(\abs{x})}$ and therefore
$\QMIP^{poly}(poly,poly) \subseteq \QMIP(poly,poly)$. This is not
necessarily an equality, because potentially, more entanglement may
be used by the provers to cheat the verifier. It is possible that
there are languages that can be proved without entanglement and can
not be proved with it.

Thus the main difference between the quantum and the classical
models is that the provers can use prior-entanglement to their
advantage, and otherwise $\QMIP=\MIP$.

The power of $\QMIP^*(poly,poly)$ is a mystery, the
provers are stronger and thus it might seem that they may prove more languages.
However the provers are also less trustworthy so there might be some
languages that the classical provers can prove but quantum provers
with entanglement can not. Thus, it is not even known that
$\QMIP^*(poly,poly)\subseteq \NEXP$ or $\NEXP \subseteq \QMIP^*(poly,poly)$.

In \cite{Julia} Kempe and Vidick expand the definition of
$\QMIP(k,1)$ to $\QMIP_{\log{n},c,s}(k,1)$, the class of languages
that have a $\QMIP$ proof with the verifiers complexity and the
message registers logarithmic in the input size. They prove that
$\NP \subseteq \QMIP^{*}_{\log{n},1,1-2^{-O(n)}}(2,1)$. This implies
that even if the provers have unlimited entanglement they can not
cheat perfectly. Recently, this result have been improved to
$1-\frac{1}{poly(n)}$ soundness. By applying the padding argument
this can be expanded to $\NEXP \subseteq
\QMIP^{*}_{poly(n),1,1-2^{-poly(n)}}(2,1)$.

\subsection{The Diamond Norm}

In this section we survey Kitaev and Watrous \cite{KitWat}
characterization of $\QIP(3)$ using the diamond norm.

\begin{deff} The \emph{Trace Norm} of an operator $A \in L(\hs{H})$ is
\[
\trnorm{A} = \max_{U \in U(\hs{H})}{\abs{\Tr(UA)}}
\]
\end{deff}

If $A$ is a normal matrix with eigenvalues $\{\lambda_i\}$ then
$\trnorm{A} = \sum_i{\abs{\lambda_i}}$. For a general $A$ it can be
checked that $\trnorm{A} = \Tr(\abs{A}) = \Tr(\sqrt{AA^{\dag}})$.
Also $\trnorm{A}=\sum_i{s_i(A)}$ where $s_1(A)\geq\cdots \geq
s_n(A)$ are the singular values of $A$. The natural generalization
of the $\trnorm{}$ to superoperators is
\begin{deff} Let $T:L(\hs{H}_1) \rightarrow L(\hs{H}_2)$ be a superoperator. The
$l_1$ norm $\norm{T}_1$ is
\[
\norm{T}_1=\max_{A:\trnorm{A}=1}\trnorm{T(A)}
\]
\end{deff}

\begin{deff} A superoperator norm $\norm{}_.$ is \emph{$f(n)$-stable}
iff for any\\ $T:L(\hs{H}_1)\rightarrow L(\hs{H}_2)$ having
$\dim(\hs{H}_1)=n$ and every $N \geq 0$ it holds that
\[
\norm{T \tensor I_N}_. \leq \norm{T \tensor I_{f(n)}}_.
\]
\end{deff}

If $f(n)=0$ we say that $\norm{}_.$ is \emph{stable}. The $l_1$ norm
is not stable. For example consider the superoperator on $L(\C^2)$
\[T(\ketbra{i}{j})= \ketbra{j}{i}
 ,(i,j=0,1)\]
On the one hand $\norm{T}_1=1$. On the other hand for
$A=\sum_{i,j}{\ketbra{i,i}{j,j}}$,
 $\trnorm{A}=2$ but $\trnorm{T \tensor I_1(A)}=4$, and so $\norm{T
\tensor I_1}_1\geq 2$.

Fortunately Kitaev \cite{kitaev2002caq} proved that $\norm{}_1$ is
$n$-stable. For any $N \geq 0$ and $n=\dim{(\hs{H}_1)}$ it holds
that $\norm{T \tensor I_N}_1 \leq \norm{T \tensor I_n}_1$. Watrous
\cite{Watrous} gave a simpler proof of that. This allows to define
the diamond norm.

\begin{deff} Let $T:L(\hs{H}_1) \rightarrow L(\hs{H}_2)$ be a
superoperator and $n=\dim{(\hs{H}_1)}$ then the diamond norm
$\dnorm{T}$ is
\[
\dnorm{T}=\norm{T \tensor I_n}_1
\]
\end{deff}

This defines a norm \cite{kitaev2002caq}. The $\dnorm{}$ is indeed
stable. Kitaev \cite{kitaev2002caq} also proved that the diamond
norm is multiplicative, i.e., $\norm{T\tensor
R}_{\diamond}=\norm{T}_{\diamond} \norm{R}_{\diamond}$. He also gave
other equivalent mathematical formulations to it.

\subsection{$\QIP(3)$ Characterization by the diamond norm}

Denote $\QIP(3,s,c)$ the class of languages with a QIP proof system
with three messages, soundness $s$ and completeness $c$. Let  $L \in
\QIP(3,s,1)$ proved to a verifier $V$. The protocol is characterized
by the unitary operators $V_1,V_2$ the verifier applies in each
round, the initial state projection $\Pi_{init}$ and the accepting
projection $\Pi_{acc}$. Denote $B_1 =
V_1\Pi_{init},B_2=\Pi_{acc}V_2$. Let $\MAP(B_1,B_2)$ denote the
maximal acceptance probability of the verifier. Kitaev and Watrous
proved that $$\MAP(B_1,B_2)=\dnorm{T}$$ where $T(X)=\Tr_{\hs{V}}(B_1
XB_2)$ giving a neat algebraic characterization of the game.

As a corollary of the above characterization and the fact that the
diamond norm is multiplicative Kitaev and Watrous showed that
$\QIP(3,s,1)$ has perfect parallel amplification.



\section{$\QMIP^*(2,1)$ and the Product Norm}

In this section we define a product operator norm and a product
superoperator norm and later prove that the maximum acceptance
probability for a given verifier in quantum one round two prover
protocol can be described in terms of it.

\subsection{The Product Norm}

\begin{deff}
For Hilbert spaces $\hs{V}_1,\hs{V}_2$ and a matrix $A \in
L(\hs{V}_1\tensor\hs{V}_2)$  the product norm of $A$ is
\[
\norm{A}_{\hs{V}_1\tensor\hs{V}_2}=\max_{U_i\in
U(\hs{V}_i)}\abs{Tr((U_1\tensor U_2)A)}
\]
\end{deff}

\begin{claim}
$\norm{}_{\hs{V}_1\tensor\hs{V}_2}$ is a norm.
\end{claim}

\begin{proof}The following things are simple.
\begin{enumerate}
\item
  $\norm{A}_{\hs{V}_1\tensor\hs{V}_2}\geq 0$.
\item
$\norm{cA}_{\hs{V}_1\tensor\hs{V}_2}=c\norm{A}_{\hs{V}_1\tensor\hs{V}_2}$.
\item
  Triangle inequality.
\item
  If $A=0$ then $\norm{A}_{\hs{V}_1\tensor\hs{V}_2}=0$.
\end{enumerate}

  We are left with showing that if $\norm{A}_{\hs{V}_1\tensor\hs{V}_2}=0$ then
  $A=0$. Assume $\norm{A}_{\hs{V}_1\tensor\hs{V}_2}=0$.
  Then $\trnorm{A}=Tr(UA)$ for some $U \in
  U(\hs{V}_1\tensor\hs{V}_2)$. The transformation $U$ can be represented as
  \[U=\sum_i{a_i(W_i\tensor V_i)}\] where $W_i \in U(\hs{V}_1) ,V_i \in
  U(\hs{V}_2)$. This is true because there is a unitary basis for
  any $L(\hs{H})$. One such possible basis is described in \cite{unibasis}.
  Thus $Tr(UA)= \sum_i{a_iTr((W_i\tensor V_i)A)}=0$ and so $\trnorm{A}=0$ and $A=0$.
\end{proof}

We notice that

\begin{equation}
\trnorm{\Tr_{\hs{V}_2}(A)} \leq
\norm{A}_{\hs{V}_1\tensor\hs{V}_2}\leq \trnorm{A}
\end{equation}

The left inequality follows from Equations (1) and (2) because\\
$\Tr((U_1 \tensor U_2)A)=\Tr(U_1\Tr_{\hs{V}_2}((I\tensor U_2)A))$.
The right inequality follows from the fact that $\max_{U_i\in
U(\hs{V}_i)}\abs{Tr((U_1\tensor U_2)A)} \leq \max_{U \in U(\hs{V}_1
\tensor \hs{V}_2)}{\abs{\Tr(UA)}}$. Those inequalities can be
strict, for example for $A$ of the form $A=\ketbra{u}{v}$. For any
such $A$, $\trnorm{A}=1$ but we will show later that for
$A=\ketbra{epr}{00}$ it holds that $\norm{A}_{\mathbb{C}^2 \tensor
\mathbb{C}^2}=\frac{1}{\sqrt{2}}$ (where $\ket{epr}=
\frac{1}{\sqrt{2}}(\ket{00}+\ket{11})$). Another example is
$A=\ketbra{00}{11}$ with the partition $\hs{V}_1=\hs{V}_2=\C^2$. On
the one hand $\trnorm{\Tr_{\hs{V}_2}(A)}=0$, but as we will show
later $\norm{A}_{\hs{V}_1 \tensor \hs{V}_2}=1$.

\subsection{The Superoperator Product Norm}

Next, we define a superoperator product norm.

\begin{deff}

For Hilbert spaces $\hs{V},\hs{V}_1,\hs{V}_2$ and superoperator
$T:L(\hs{V})\rightarrow L(\hs{V}_1\tensor\hs{V}_2)$ the
superoperator product norm is
\[
\norm{T}_{\hs{V}_1\tensor\hs{V}_2,tr}=\max_{\trnorm{A}=1}\norm{T(A)}_{\hs{V}_1\tensor\hs{V}_2}
\]

\end{deff}

It is easy to check that this is a norm and that
$\norm{I}_{\hs{V}_1\tensor\hs{V}_2,tr}=1$. Also, it follows from
Equation (3) that $\norm{T}_{\hs{V}_1\tensor\hs{V}_2,tr} \leq
\dnorm{T}$. A useful fact is:
\begin{claim}
\[
\norm{T}_{\hs{V}_1\tensor\hs{V}_2,tr}= \max_{\ket{u},\ket{v} \in
\hs{V}}\norm{T(\ketbra{u}{v})}_{\hs{V}_1\tensor\hs{V}_2}
\]
\end{claim}

\begin{proof}
Any $A$ satisfying $\trnorm{A}=1$ has a singular value decomposition
$A=\sum_i{s_i\ketbra{u_i}{v_i}}$ for $s_i \geq 0$ and
$\sum_i{s_i}=1$. Thus

\begin{eqnarray*}
\norm{T(A)}_{\hs{V}_1\tensor\hs{V}_2}
&=&\norm{T(\sum_i{s_i\ketbra{u_i}{v_i}})}_{\hs{V}_1\tensor\hs{V}_2}\\
&\leq&
\sum_i{s_i\norm{T(\ketbra{u_i}{v_i})}_{\hs{V}_1\tensor\hs{V}_2}}\\
&\leq& \max_i{\norm{T(\ketbra{u_i}{v_i})}_{\hs{V}_1\tensor\hs{V}_2}}
\end{eqnarray*}

Thus the maximum is always achieved on some rank one matrix
$\ketbra{u}{v}$.
\end{proof}

\begin{claim} For any two superoperators
$T:L(\hs{H}_1)\rightarrow L(\hs{V}_1 \tensor \hs{V}_2)$ and
$R:L(\hs{H}_2) \rightarrow L(\hs{W}_1 \tensor \hs{W}_2)$ it holds
that

\[\norm{T \tensor R}_{(\hs{V}_1 \tensor\hs{W}_1) \tensor (\hs{V}_2
\tensor\hs{W}_2),tr}\geq \norm{T}_{\hs{V}_1 \tensor \hs{V}_2,tr}
\cdot \norm{R}_{\hs{W}_1 \tensor \hs{W}_2,tr}\]
\end{claim}

\begin{proof}

$$\norm{T \tensor R}_{(\hs{V}_1 \tensor\hs{W}_1) \tensor (\hs{V}_2
\tensor\hs{W}_2),tr} = \max_{\trnorm{X}=1}\norm{(T \tensor
R)(X)}_{(\hs{V}_1 \tensor\hs{W}_1) \tensor (\hs{V}_2
\tensor\hs{W}_2)}$$

Let us look at the special case where $X \in L(\hs{H}_1 \tensor
\hs{H}_2)$ is product, $X=A\tensor B$ for some
 $A\in L(\hs{H}_1)$ and $B \in L(\hs{H}_2)$.

\begin{eqnarray*}
\norm{T \tensor R}_{(\hs{V}_1 \tensor\hs{W}_1) \tensor (\hs{V}_2
\tensor\hs{W}_2),tr}
 &\geq& \max_{\trnorm{A}=\trnorm{B}=1}\norm{(T \tensor R)(A \tensor
B)}_{(\hs{V}_1 \tensor\hs{W}_1) \tensor (\hs{V}_2 \tensor\hs{W}_2)}\\
&=& \max_{\trnorm{A}=\trnorm{B}=1}\norm{T(A) \tensor
R(B)}_{(\hs{V}_1
\tensor\hs{W}_1) \tensor (\hs{V}_2 \tensor\hs{W}_2)}\\
&=&\max_{\trnorm{A}=\trnorm{B}=1,U_1,U_2}\Tr((U_1 \tensor U_2)(T(A)
\tensor R(B)))
\end{eqnarray*}

for unitaries $U_1\in U(\hs{V}_1 \tensor\hs{W}_1)$ and $U_2 \in
U(\hs{V}_2 \tensor\hs{W}_2)$. We again look at the special case
where $U_1$ and $U_2$ are also products of unitaries $U_1 = V_1
\tensor W_1$ and $U_2 = V_2 \tensor W_2$ for $V_1 \in U(\hs{V}_1),
W_1 \in U(\hs{W}_1),V_2 \in U(\hs{V}_2), W_2 \in U(\hs{W}_2)$.
Then \\

$\begin{array}{l} \norm{T \tensor R}_{(\hs{V}_1 \tensor\hs{W}_1)
\tensor (\hs{V}_2
\tensor\hs{W}_2),tr}\\
\geq
 \max_{\trnorm{A}=\trnorm{B}=1,V_1,V_2,W_1,W_2}\Tr((V_1 \tensor W_1
\tensor V_2 \tensor W_2)(T(A) \tensor R(B)))\\
= \max_{\trnorm{A}=\trnorm{B}=1,V_1,V_2,W_1,W_2}\Tr((V_1 \tensor
V_2)T(A))
\cdot \Tr((W_1 \tensor W_2)R(B))\\
=  \norm{T}_{\hs{V}_1 \tensor \hs{V}_2,tr} \cdot \norm{R}_{\hs{W}_1
\tensor \hs{W}_2,tr}
\end{array}$

\end{proof}

In particular it follows from above that $$\norm{T}_{\hs{V}_1
\tensor \hs{V}_2,tr} \leq \norm{T \tensor I_{\hs{W}_1 \tensor
\hs{W}_2 }}_{(\hs{V}_1 \tensor \hs{W}_1)\tensor(\hs{V}_2 \tensor
\hs{W}_2),tr}$$

Next we expand the definition of stability to the superoperator
product norm. We do this by adding to each register of the original
partition $\hs{V}_1,\hs{V}_2$ an additional register $\C^N$ and
applying the superoperator $T\tensor I_{N}\tensor I_{N}$ with the
identity operator over the new registers.

\begin{definition}
A $\norm{}_{\hs{V}_1 \tensor \hs{V}_2,tr}$ is $f(n)$-stable iff for
any $T:L(\hs{H})\rightarrow L(\hs{V}_1 \tensor \hs{V}_2)$ having
$\dim(\hs{H})=n$ and every $N \geq 0$ it holds that
\[
\norm{T \tensor I_{N^2}}_{(\hs{V}_1\tensor \mathbb{C}^N) \tensor
(\hs{V}_2\tensor \mathbb{C}^N),tr } \leq \norm{T \tensor
I_{f(n)^2}}_{(\hs{V}_1\tensor \mathbb{C}^{f(n)}) \tensor
(\hs{V}_2\tensor \mathbb{C}^{f(n)}),tr}
\]
\end{definition}

The $\norm{}_{\hs{V}_1 \tensor \hs{V}_2,tr}$ norm is not 0 stable.
Consider the superoperator \\$T:L(\mathbb{C}^4)\rightarrow
L(\C^2 \tensor \C^2)$ that is defined by \\
$T(\ketbra{i,j}{k,m})=\ketbra{k,m}{i,j}$. Then
$\norm{T}_{\mathbb{C}^2 \tensor \mathbb{C}^2,tr} \leq \norm{T}_1
=1$. On the other hand, $\norm{T \tensor I_4}_{(\mathbb{C}^2 \tensor
\mathbb{C}^2) \tensor (\mathbb{C}^2 \tensor \mathbb{C}^2),tr}=4$. To
see that use $A=\sum_{i,j,k,m}{\ketbra{i,j,i,j}{k,m,k,m}}$. It is
easy to check that $\trnorm{A}=4$, and that by
$U\ket{i,k}=\ket{k,i}$ we have \\$(U \tensor U)(T\tensor I_4)
(A)=\sum_{i,j,k,m}{\ketbra{i,k,j,m}{i,k,j,m}}=I_{16}$ and so\\
$\norm{T \tensor I_4}_{(\mathbb{C}^2 \tensor \mathbb{C}^2) \tensor
(\mathbb{C}^2 \tensor \mathbb{C}^2),tr}\geq 4$. Altogether
\\$\norm{T \tensor I_4}_{(\mathbb{C}^2 \tensor \mathbb{C}^2) \tensor
(\mathbb{C}^2 \tensor \mathbb{C}^2),tr} \leq\norm{T}_{\diamond}=4 $.

\subsection{$\QMIP^*(2,1)$}

In this section we focus on $\QMIP^*(2,1)$. The protocol is applied
on the registers $\hs{V} \tensor \hs{M}_1 \tensor \hs{M}_2 \tensor
\hs{P}_1 \tensor \hs{P}_2$ where $\hs{V}$ is the verifier's private
register. $\hs{M}_1$, $\hs{M}_2$ are the registers passed between
$V$ and $P_1$, $P_2$ respectively. $\hs{P}_1$, $\hs{P}_2$ are the
private registers of the provers. The initial quantum state is some
$\ket{\psi}$ of an arbitrary length chosen as part of the prover
strategy.

The protocol proceeds as follows:
\begin{enumerate}
  \item
  The verifier applies a measurement defined by
  $\Pi_{init}=\ketbra{0}{0}$ on \\
  $\hs{V} \tensor \hs{M}_1\tensor \hs{M}_2$. If the outcome is not $\ket{0}$ he rejects.
  This step checks the initial state.
  \item
  The verifier applies a unitary transformation $V_1$ on
  $\hs{V} \tensor \hs{M}_1\tensor \hs{M}_2$. This prepares the
  questions to the two provers.
  \item
  Prover $i$ applies a unitary $U_i$ on $\hs{M}_i\tensor \hs{P}_i$.
  \item
  The verifier applies a unitary $V_2$ on $\hs{V} \tensor \hs{M}_1\tensor
  \hs{M}_2$, followed by a measurement defined by $\Pi_{acc}=\ket{0}\bra{0}$
  on the first qubit of $\hs{V}$ and accepts iff the outcome is
  $\ket{0}$.
\end{enumerate}

If the provers are successful in convincing the verifier the final
(unnormalized) state of the system is thus $$((\Pi_{acc}V_2)\tensor
I_{\hs{P}_1 \tensor \hs{P}_2})(I_\hs{V} \tensor U_1 \tensor
U_2)((V_1\Pi_{init})\tensor I_{\hs{P}_1 \tensor
\hs{P}_2})\ket{\psi}$$

\subsection{Acceptance Probability for a Given Verifier}

Let $V$ be a verifier. $V$'s strategy is defined by
$B_1=V_1\Pi_{init}$ and $B_2=\Pi_{acc}V_2$. Let $\MAP(B_1,B_2)$
denote the maximum acceptance probability of $V$, when $V$ plays
with the optimal provers. I.e.,

\begin{eqnarray}
\MAP(B_1,B_2)= \max_{U_i\in U(\hs{M}_i\tensor\hs{P}_i), \ket{\psi}}
\abs{(B_2\tensor I_{\hs{P}_1 \tensor \hs{P}_2})(I_\hs{V} \tensor U_1
\tensor U_2)(B_1\tensor I_{\hs{P}_1 \tensor \hs{P}_2})\ket{\psi}}^2
\end{eqnarray}

We now relate $MAP(B_1,B_2)$ to the superoperator product norm. We
claim that:

\begin{theorem}
$\MAP(B_1,B_2)= \norm{T\tensor I_{\hs{P}_1 \tensor
\hs{P}_2}}_{(\hs{M}_1\tensor\hs{P}_1)\tensor(\hs{M}_2\tensor\hs{P}_2),tr}^2
$\\

where $T:L(\hs{V}\tensor \hs{M}_1 \tensor\hs{M}_2 )\rightarrow L(\hs{M}_1 \tensor\hs{M}_2)$
is defined by $T(X)=Tr_{\hs{V}}(B_1XB_2)$.
\end{theorem}

\begin{proof} Denote $\hs{P}=\hs{P}_1 \tensor \hs{P}_2$. We start
with Equation (4).
\[
\sqrt{\MAP(B_1,B_2)}=\max_{U_1,U_2,\psi}\abs{(B_2\tensor
I_\hs{P})(I_\hs{V} \tensor U_1 \tensor U_2)(B_1\tensor
I_\hs{P})\ket{\psi}}
\]
Since we maximize over the unit vector $\ket{\psi}$ we can replace
the vector norm with the operator norm
\[
\sqrt{\MAP(B_1,B_2)}=\max_{U_1,U_2}\norm{(B_2\tensor
I_\hs{P})(I_\hs{V} \tensor U_1 \tensor U_2)(B_1\tensor I_\hs{P})}
\]
The operator norm of the matrix is the largest singular value, and
so
\[
\sqrt{\MAP(B_1,B_2)}=\max_{U_1,U_2,v,u}\abs{\bra{v}(B_2\tensor
I_\hs{P})(I_\hs{V} \tensor U_1 \tensor U_2)(B_1\tensor
I_\hs{P})\ket{u}}
\]

Since this is a scalar number we can insert trace
\begin{eqnarray*}
\sqrt{\MAP(B_1,B_2)} &=&
\max_{U_1,U_2,v,u}\abs{\Tr(\bra{v}(B_2\tensor I_\hs{P})(I_\hs{V}
\tensor U_1 \tensor U_2)(B_1\tensor I_\hs{P})\ket{u})}\\
&=& \max_{U_1,U_2,v,u}\abs{\Tr((I_\hs{V} \tensor U_1 \tensor
U_2)(B_1\tensor I_\hs{P})\ketbra{u}{v}(B_2\tensor I_\hs{P}))}
\end{eqnarray*}

By Equation (1)
\[
\sqrt{\MAP(B_1,B_2)}=\max_{U_1,U_2,v,u}\abs{\Tr(\Tr_{\hs{V}}((I_{\hs{V}}
\tensor U_1 \tensor U_2)(B_1\tensor
I_\hs{P})\ketbra{u}{v}(B_2\tensor I_\hs{P})))}
\]

By Equation (2) we can carry the operators that do not affect
$\hs{V}$ out, use the definition of $T$ and then use Claim 2.
\begin{eqnarray*}
\sqrt{\MAP(B_1,B_2)} &=& \max_{U_1,U_2,u,v}\abs{\Tr((U_1 \tensor
U_2)Tr_{\hs{V}}((B_1\tensor I_\hs{P})\ketbra{u}{v}(B_2\tensor
I_\hs{P})))}\\
&=& \max_{U_1,U_2,u,v}\abs{\Tr((U_1 \tensor U_2)(T\tensor
I_{\hs{P}})(\ketbra{u}{v}))}\\
&=& \norm{T\tensor
I_{\hs{P}}}_{(\hs{M}_1\tensor\hs{P}_1)\tensor(\hs{M}_2\tensor\hs{P}_2),tr}
\end{eqnarray*}
\end{proof}

Let us notice that this proof is almost identical to the proof of
$\QIP(3)$ characterization by Kitaev and Watrous \cite{KitWat}. The
main difference is that here we have a product norm instead of the
trace norm as a target. This is because the initial state of the
provers in $\QMIP^{*}(2,1)$ can be viewed as the first message and
so we actually have three messages instead of two.

\section{Product Norm of Rank 1 Matrices}

We start with a useful bound on $\trnorm{BC}$ and use it to show
what is the product norm for rank 1 matrices.

\begin{lemm}
\label{lem:hj}
Fix arbitrary matrices  $B$ and $C$ with
   $s_1(B)\geq \cdots \geq s_n(B) \geq 0$ the singular values of $B$,
   and $s_1(C)\geq \cdots \geq s_n(C) \geq 0$ the singular values of $C$. Then
  \[
  \trnorm{BC} \leq \sum_i{s_i(B)s_i(C)}
  \]
\end{lemm}


The above claim appears in \cite{horn1994tma} (page 182, Exercise
4). Notice also that this is tight for normal commuting matrices $B$
and $C$.

With that we prove:

\begin{theorem}Let $A$ be a rank $1$ matrix over $\hs{V}_1\tensor
\hs{V}_2$. Thus $A=\ketbra{u}{v}$ for some $u,v \in \hs{V}_1\tensor
\hs{V}_2$. Suppose the Schmidt decomposition of $u$ is
$\ket{u}=\sum_i{\alpha_i\ket{x_i}\tensor\ket{y_i}}$, and of $v$ is
$\ket{v}=\sum_i{\beta_i\ket{w_i}\tensor\ket{z_i}}$ with
$\alpha_i,\beta_i\geq 0$ sorted in descending order. Then
\[
    \norm{A}_{\hs{V}_1\tensor\hs{V}_2}=\sum_i{\alpha_i\beta_i}
\]
\end{theorem}

\begin{proof}
We can assume without loss of generality that
$\ket{x_i}=\ket{y_i}=\ket{w_i}=\ket{z_i}=\ket{i}$ because
$\norm{A}_{\hs{V}_1\tensor\hs{V}_1}=\norm{(U_1\tensor
U_2)A(V_1\tensor V_2)}_{\hs{V}_1\tensor\hs{V}_1}$ for any unitaries
$U_1,V_1 \in U(\hs{V}_1)$ and $U_2,V_2 \in U(\hs{V}_2)$. Thus

\[A=\ketbra{u}{v}=\sum_{i,j}{\alpha_i\beta_j\ketbra{i,i}{j,j}}\]
and
\begin{eqnarray*}
\Tr{((U_1\tensor
U_2)A)}&=&\sum_{i,j}{\alpha_i\beta_j\bra{j}U_1\ket{i}\bra{j}U_2\ket{i}}\\
&=& \sum_{i,j}{\alpha_i(U_1)_{j,i} \cdot \beta_j(U_2)_{j,i}}
\end{eqnarray*}
We can look at this sum of products as a standard matrix inner
product. Let us denote the matrices $C$ and $B$ as follows,
$C_{j,i}=\alpha_i(U_1)_{j,i}$ and $B_{j,i}=\beta_j(U_2)_{j,i}$. Then
$$\Tr{((U_1\tensor U_2)A)}=\sum_{i,j}{B_{i,j}C_{i,j}}=\Tr(B^{t}C)$$
By Lemma 4.1, $\abs{\Tr(B^{t}C)}\leq \trnorm{B^{t}C}\leq
\sum_i{\alpha_i\beta_i}$, because $C=U_1 \diag(\alpha_1,\ldots
\alpha_n)$, $B=\diag(\beta_1,\ldots \beta_n)U_2$, and so
$s_i(C)=\alpha_i$ and $s_i(B)=\beta_i$. Finally, this upper bound
can be achieved by $U_1=U_2=I$.
\end{proof}

\section{Directions for Further Research}

We can \emph{not} prove that the product norm stabilizes. However we
would like to check what such a result would give.

\begin{hypothesis}$\norm{}_{\hs{V}_1 \tensor \hs{V}_2,tr}$ is $poly(n)$-stable.
\end{hypothesis}

\begin{claim}
Under hypothesis 1 $\QMIP^*(2,1)\subseteq \NEXP = \MIP(2,1)$.
\end{claim}

\begin{proof}
Let $L \in \QMIP^*(2,1)$. Consider a verifier $V$ for $L$. By
Theorem 5.1 the maximum acceptance probability of $V$ is
$$\MAP(B_1,B_2)=\norm{T\tensor I_{\hs{P}_1 \tensor \hs{P}_2}}_{(\hs{M}_1\tensor\hs{P}_1)\tensor(\hs{M}_2\tensor\hs{P}_2),tr}^2$$
for $B_1,B_2$ and $T$ defined as before. It follows from Definitions
6,7 and Claim 2 that
$$\MAP(B_1,B_2)= Tr((U_1 \tensor U_2)(T \tensor I_{\hs{P}_1 \tensor \hs{P}_2})(\ketbra{u}{v}))$$
for some $U_1 \in U(\hs{M}_1 \tensor \hs{P}_1)$, $U_2 \in U(\hs{M}_2
\tensor \hs{P}_2)$ and $\ket{u},\ket{v} \in \hs{V} \tensor \hs{M}_1
\tensor \hs{M}_2 \tensor \hs{P}_1 \tensor \hs{P}_2$. Under the
hypothesis we can fix such  $U_1,U_2$ and $\ket{u},\ket{v}$ that
live in the world of $poly(\abs{x})$ qubits. Consider the prover
strategy $U_1 \tensor U_2$ with the initial state $\ket{u}$. This
strategy uses only $poly(\abs{x})$ entangled qubits in the initial
state and is optimal. Thus $\QMIP^*(2,1) \subseteq
\QMIP^{poly}(2,1)$ and we already mentioned that Kobayashi and
Matsumoto proved in \cite{QMIP} that
$\QMIP^{poly}(2,1)\subseteq\NEXP$.
\end{proof}

Another hypothesis is the following. It is not known if there exists
an efficient Turing machine for approximating the $\dnorm{}$.
However Kitaev and Watrous proved in \cite{KitWat} that $\QIP
\subseteq \EXP$ by showing a reduction from distinguishing between
the case of $\MAP(B_1,B_2)=1$ and $\MAP(B_1,B_2) \leq \frac{1}{2}$
to a semidefinite programming problem of an exponential size (in the
number of qubits).

\begin{hypothesis}
For $T:L(\hs{H})\rightarrow L(\hs{V}_1 \tensor \hs{V}_2)$ there
exists a Turing machine that approximates $\norm{T}_{\hs{V}_1
\tensor \hs{V}_2,tr}$ in $poly(\dim(\hs{H})+\dim(\hs{V}_1 \tensor
\hs{V}_2))$ time.
\end{hypothesis}

\begin{claim}
If both hypotheses are true then $\QMIP^*(2,1) \subseteq \EXP$.
\end{claim}

\begin{proof}
Let $L \in \QMIP^*(2,1)$. Hypothesis 1 implies that $L$ has a protocol \\
$\la V,P_1,P_2 \ra$ with maximum acceptance probability
$$\norm{T\tensor I_{\hs{P}_1 \tensor
\hs{P}_2}}_{(\hs{M}_1\tensor\hs{P}_1)\tensor(\hs{M}_2\tensor\hs{P}_2),tr}^2$$
for $T$ defined as previously and
$\dim(\hs{M}_1\tensor\hs{P}_1\tensor
\hs{M}_2\tensor\hs{P}_2)=2^{poly(\abs{x})}$. Hypothesis 2 implies
that a there is a Turing machine that approximates the maximum
acceptance probability and decides if $x \in L$ in
$poly(2^{poly(\abs{x})})$ time.
\end{proof}

\section*{Acknowledgements}

We thank Julia Kempe, Ashwin Nayak and Oded Regev for interesting
discussions on the subject. We thank Zeph Landua for helping us
prove Lemma \ref{lem:hj}, Ashwin Nayak for an alternative proof and
Oded Regev for refereeing us to the place where it is proven in
\cite{horn1994tma}.

\bibliographystyle{plain}
\bibliography{bibl}

\end{document}